\documentstyle[12pt]{article}
\newcommand{\be}{\begin{eqnarray}}
\newcommand{\ee}{\end{eqnarray}}
\newcommand{\ben}{\begin{eqnarray*}}
\newcommand{\een}{\end{eqnarray*}}
\newcommand{\bc}{\begin{center}}
\newcommand{\ec}{\end{center}}

\newcommand{\as}{\alpha_s}
\newcommand{\sinfi}[1]{\sin\varphi({#1})}
\newcommand{\cosfi}[1]{\cos\varphi({#1})}
\newcommand{\ds}{\displaystyle}
\newcommand{\Vqp}{V(|q-p|)}
\newcommand{\Vpq}{V(|p-q|)}

\def\thebibliography#1{\section*{References}\list
 {\arabic{enumi}.}{\settowidth\labelwidth{[#1]}\leftmargin\labelwidth
 \advance\leftmargin\labelsep
 \usecounter{enumi}}
 \def\newblock{\hskip .11em plus .33em minus -.07em}
 \sloppy
 \sfcode`\.=1000\relax}

\begin{document}
\bc{\Large  Pion in the quark model.}\\
{\bf Obid Juraev} \\
{\it Abdus Salam International Centre for theoretical Physics, Trieste
34100, Italy}\\
e-mail : juraev@ictp.trieste.it\\
{\bf Tulkun Nasyrov} \\
{\it Institute of Nuclear Physics, 702132 Tashkent, Ulugbek, Uzbekistan}\\
e-mail:   nasirt@suninp.tashkent.su\\

\ec
\large 
{ 1. Introduction.

In last years theoretical analyses of light and heavy quarkonia in
relativistic models draws the great attention. It is known, that 
criterion for the testing of hadronization models in
Quantum chromodynamics (QCD) is the leptonic constant of pion decay 
as scale for the dynamics of hadrons was considered in work [2].
On the one hand of constant decay acts as  phoundamental  parameter
in local effective chiral Lagrangians, on the other hand it is bound with
nonlocal potential model through wave function of bound state in zero.
In [3,4], where the effective hamiltanian QCD with simultaneously fourquark
interaction is used, the experimental value $F_{\pi}$ is  failed
to be reproduce.

	In this paper it  will be shown that self-consistent defination of
leptonic decay constant which
simultaneously satisfies Goldstone theorem and the normalization condition are
gives the close value to the experimental data.

Using the "minimum" quantanization method [5],  based on solving the
Gauss equation and on the principle of gauge invariant  the action
corresponding to chromostatic  approximation [5] should be written as
follows:
\be
S(q,\bar q) = \int d^{4}xd^{4}y\bar q(x) G^{-1}_{0} q(y) 
+ \frac{1}{2}
\int d^{4} x_{1}d^{4}y_{1}d^{4} x_{2} d^{4} y_{2} 
\nonumber \\ q_{D_{1}}(y_{1})
\bar q_{H_{1}}(x_{1})          
K_{D_{1}H_{1}/D_{2}H_{2}}(x_{1}y_{1}/x_{2}y_{2})
q_{H_{2}}(y_{2}) q_{D_{2}}(x_{2})
\ee
where $G^{-1}_{0}(x,y) = (i \rlap/ \partial - \widehat m_{0}) \delta^{4}(x-y)$, 
$$K_{D_{1}H_{1}/D_{2}H_{2}}(x_{1}y_{1}/x_{2}y_{2})=
\delta_{i_{1}j_{2}}\delta_{i_{2}j_{1}}
(\gamma_{0})_{\alpha_{1}\beta_{2}}(\gamma_{0})_{\alpha_{2}\beta_{1}}$$ 
$$
\times\left( \frac{\ds{\lambda_{a}}}{\ds{2}} \right)^{A_{1}B_{2}}
\left( \frac{\ds \lambda_{a}}{\ds 2} \right)^{A_{2}B_{1}}
\delta^{4}(x_{1}-y_{2})
\delta^{4}(x_{2}-y_{1})
\delta(x^{0}_{1}-x^{0}_{2}) V_{s}(\vec x_{1} - \vec x_{2}) 
$$
where $D_{k}$ and  $H_{k}$ $(k=1,2)$ -- are complex of colour, flavour 
and Lorentz indices $(A_{k},i_{k},\alpha_{k})$ and 
$(B_{k},j_{k},\beta_{k})$ respectively, $V_{s}$ -- inte\-rac\-ti\-on potential. 
Taking into account factorize notions
$\{N_{c}\}$ and $\{N^{*}_{c}\}$ colour group
\ben
\{N_{c}\} \otimes \{N^{*}_{c}\} = \{1\} \oplus \{N^{2}_{c}-1\}
\een
 the potential $K$ can be decomposed into the colour singlet and
the $N^{2}_{c} -1$ -- plet components through the projective operators
$P_{1}$ and  $P_{N^{2}_{c} -1}$ :
\ben
\sum^{N^{2}_{c} -1}_{a=1}
\left( \frac{\lambda_{a}}{2}  \right)^{A_{1}B_{2}}
\left( \frac{\lambda_{a}}{2}  \right)^{A_{2}B_{1}}
= \frac{1}{2} \left( \frac{N^{2}_{c} -1}{N_{c}} P_{1} -
\frac{1}{N_{c}}P_{N^{2}_{c} -1} \right)
\een
Since we shall investigate colourless objects, further, we
shall consider only singlet chanel, i.e. $K=K_{1}$ ,   \\
where
$P^{A_{1}B_{2}/A_{2}B_{1}}_{1}=\delta^{A_{1}B_{1}}\delta^{A_{2}B_{2}}$

The effective covariant relativistic covariant action for singlet
cha\-n\-nel is written as:
\be
S_{íää}=\int d^{4}x \bar q(x) (i \rlap/\partial - \widehat m_{0}) q(x) -
\frac{1}{2} \int d^{4}x d^{4}y
q_{\beta_{2}}(y)\bar q_{\alpha_{1}}(x) \times \nonumber\\
\left[ K_{1}(x-y|\frac{x-y}{2}
\right]_{\alpha_{1},\beta_{1},\alpha_{2},\beta_{2}}
q_{\beta_{1}}(x)\bar q_{\alpha_{2}}(x)
\ee
with interaction kernel \\
$K_{1}(x-y|X)_{\alpha_{1},\beta_{1}/
\alpha_{2}\beta_{2}}= \rlap/\eta_{\alpha_{1}\beta_{1}}
[V(z^{\perp}\delta(z\cdot\eta)]
\rlap/\eta_{\alpha_{2}\beta_{2}}$
quantization axis is chosen
$\eta_{\mu} \sim \frac{\ds 1}{\ds i} \frac{\ds \partial}{\ds\partial X_{\mu}}$,
$\eta_{\mu}$ -- single time-like fourth-vector of time axis.

\bc 2. Schwinger-Dyson and Bete-Salpeter equations.\ec

The covariant-effective action (2) leads to the following equations:

SD equation over function $\varphi(p)$  and energy $E(p)$ of quark having
nonzero mass. The SD-equation  define one-particle energy of flavour quark.
\be
\begin{array}{lcl}
E(p)\sinfi{p} & = &m^{0}+
\ds{\frac{1}{2}} \ds{\int} (dq) \Vqp \sinfi{q} \\
E(p)\cosfi{p}& = &p+
\ds{\frac{1}{2}} \ds{\int} (dq) \Vqp \hat p \hat q \cosfi{q}
\end{array}
\ee
where $(dq)=\frac{\ds d^3q}{\ds (2\pi)^3}$, $p=|\vec p|$, 
${\hat p} =\frac{\ds \vec p}{\ds p}$
The bound state of quark-antiquark system is described by BS equation
\be
\begin{array}{ccl}
ML_{1}(p)&=&E_{T}(p)L_{2}(p) \\
&+&\ds{\int} (dq) \Vpq \left\{c^{+}_{p}c^{+}_{q}+
\hat p \hat q s^{+}_{p}s^{+}_{q}
\right\}L_{2}(q)                              \\
ML_{2}(p)&=&E_{T}(p)L_{1}(p) \\
&+&\ds{\int} (dq) \Vpq \left\{s^{-}_{p}s^{-}_{q} +
 \hat p \hat q c^{+}_{p}c^{+}_{q}
\right\}L_{1}(q)
\end{array}
\ee
where $M$ -- eigenvalue, $L_{1,2}(p)$ -- eigenfunctions,\\
$s^{\mp}_{p}=\sin \left[ \frac{\ds \varphi_1(p)\mp\varphi_2(p)}{\ds 2}\right]$, 
$c^{\mp}_{p}=\cos \left[ \frac{\ds \varphi_1(p)\mp\varphi_2(p)}{\ds 2}\right]$, \\
$E_{T}(p)=E_1(p)+E_2(p)$, $\varphi_{1,2}(p)$ -- energy and wave
function (the solution of the (4) equation) of particles (quark) 1 and 2.

Eigenfunctions $L_{1,2}$ define wave function of bound state in rest frame 
$\psi(q)=\gamma_5[L_1(q)+\gamma_{0}L_2(q)]$, \\
the normalization condition is
\be
\frac{4N_c}{M} \int (dq) L_1(q)L_2(q) = 1
\ee

In the chiral limit, when current quarks  masses are equal to zero and
$L_2=0$, the equation (4) reproduces SD equation (3), which corresponds
to the Goldstone theorem for consistent pseudoscalar meson with nonzero
 mass [5]. Corresponding (goldstone solution) solution:
\be
L_{1}(q) = \frac{1}{F_{\pi}} \sin\left[\frac{\varphi_{1}(q)+
\varphi_{2}(q)}{2} \right]
\ee
$\varphi_{1,2}(p)$ -- the solution of the SD equation in zero current quark 
mass limit.

Thus, the solution of BS equation (4) must be satisfy the normalization 
condition (5) and (in the chiral limit) the Goldstone theorem (6).

\bc 3.The interaction potential of $q \bar q$ -- system.\ec

The potential phenomenology of spectroscopy of quarkonia [3,4] in
the meaning of the real interaction, uses in general, the sum coulomb
and the increasing potential. In the increasing potential is used the lattice 
calculation for description of the heavy quarkonia.
The defination of the form $q \bar q$ -- potential is one of the actual
problem of the meson spectroscopy. We chose the potential
in the form as sum of oscillator and Coulomb terms.
\be
V(p)= - \frac{4}{3} \left[\frac{4\pi \alpha_s}{p^2} +
(2\pi)^3 V^{3}_{0}\triangle_{p}\delta^3(p)\right]
\ee

One particle SD equation (3) for quark in dimensionless units takes a form
\be
\frac{d^2\varphi(p)}{dp^2}+\frac{2}{p} \frac{d\varphi(p)}{dp} +
\frac{\sin 2\varphi(p)}{p^2}-2p\sinfi{p}+2m^0\cosfi{p} \nonumber\\
+\frac{2\as}{3\pi}\int dq \frac{q}{p}
\left\{{\cal R}_1(p,q)
\cosfi{p} -
{\cal R}_2(p,q)
\sinfi{p}\right\} = 0
\ee
\be
E(p)&=&p\cosfi{p}+m^0\sinfi{p}-\frac{1}{2}[\varphi'(p)]^2 \nonumber\\
& &+\frac{2\alpha_s}{3\pi} \int dq \frac{q}{p}
\left\{{\cal R}_1(p,q)
\sinfi{p} -
{\cal R}_2(p,q)
\cosfi{p}\right\}
\ee
where
\ben
{\cal R}_1(p,q)&=&
\ln{\left| {p+q \over p-q} \right|}
\left[\sinfi{q}-\frac{m^0}{r(q)}\right], \\
{\cal R}_2(p,q)&=& \left({p^2 +q^2 \over 2pq}
\ln{\left| {p+q \over p-q} \right|}-1\right)
\left[\cosfi{q}-\frac{q}{r(q)}\right],
\een
$r(q)=\sqrt{q^2+(m^0)^2} $

The BS equation with potential (7) in dimensionless
units takes a form of the integral-differential equation
\be
\begin{array}{c}
{\widehat D}^{(-)}L_1(p)+\ds{\frac{\alpha_s}{\pi}}
{\widehat I}^{(-)} L_1(q) =  M L_2(p) \vspace{0.4cm} \\
{\widehat D}^{(+)} L_2(p)+\ds{\frac{\alpha_s}{\pi}}
{\widehat I}^{(+)} L_2(q) =  M L_1(p)
\end{array}
\ee
where
\ben
& &{\widehat D}^{(\mp)} L(p) = -\left[ {d^2 \over dp^2} + \frac{2}{p} {d\over dp}
-E_{T}(p) + (\varphi'_{\mp})^2 + \frac{2}{p} (s^{(\mp)}_p)^2 \right] L(p) \\
& &{\widehat I}^{(\mp)} L(q) = \int dq \frac{q}{p} \left\{
{\cal R}_1(p,q) c^{(\mp)}_{q} c^{(\mp)}_{p} +
{\cal R}_1(p,q) s^{(\mp)}_{q} s^{(\mp)}_{p} \right\}L(q)
\een

The main equations (8), (9), (10) are solved by the numerical method which
is given in the papers [6,7]. On fig.1 the solution of BS eqution (10) 
for the different values of the coulomb potential constant ($\alpha_s$) 
is given, 
constituent quark mass $m^0 = m^0_1= m^0_2$ and $m^0= 0.021(4V^3_0/3)^{1/3}$.

\bc 4.The pion constant decay.\ec

The self-consistent defination of the pion leptonic decay constant $F_\pi$ 
which satisfies to the Goldstone theorem (in the chiral limit) and the 
normalization 
condition simultaneously gives the close value to experimental data. According
to the  paper [8] $F_\pi$ is defined as
\be
F_{\pi}= \frac{4N_c}{M_\pi} \int (dq) L_2(q) \sin \left[\frac{
\varphi_1(q)+\varphi_2(q)}{2}\right]
\ee

At small current quark masses BS-equation haves the approximate solution
\be L_{1}(p) \approx \frac{\sinfi{p}}{F_\pi}\ee
\be L_2(p) \approx \frac{m^0}{M_{\pi}F_\pi}\ee
where $m^0=m^0_u=m^0_d$. If (13) substitute for (12), then 
\ben M^2_{\pi}F^2_{\pi}= 2m^0\left( 2N_c\int(dq)\sinfi{q}\right)\een
is obtained the expression in brackets is considered in paper [9], this
value is called the quark condensate $-<\bar q q>$. Then it can be written
\ben M^2_{\pi}F^2_{\pi}= - 2m^0 <\bar q q> , \een
well-known algebra current correlation.

It is interesting to note that the approximate solutions (12) and (13),
in the case of small current quarks of mass $m^0$, allows to make the 
qualitative estimation of BS equation (10) solution which is founded Fig.1. 
Let us consider the calculated wave functions $L_1(p)$ and $L_2(p)$ in the 
limit $p\rightarrow 0$. From Fig.1 you can see, that at $\alpha_s=0.56$
\ben
L_1(0)=5.826 \left(\frac{4}{3}V^3_0\right)^{-1/3}, \\
L_2(0)=0.4566\left(\frac{4}{3}V^3_0\right)^{-1/3}
\een

Since $\varphi(0)=\pi/2$ from (12) and (13) we have
\be M_{\pi}=m^0\frac{L_1(0)}{L_2(0)} \qquad \mbox{¨«¨} \qquad
m^0=M_{\pi}\frac{L_2(0)}{L_1(0)}\ee
\be F_{\pi}=\frac{\left(\frac{4}{3}V^3_0\right)^{1/3}}{L_1(0)}\ee

The formula (14) gives the possibility to calculate the current
quark mass independence on the oscillator potential parameter it is 
sufficiently to know the wave function in zero.
If take the experimental value of pion mass  $M_{\pi}=140 MeV$, then
the $u,\ d$ quark mass  $m^0=11\ MeV$ will be obtained.

The form (15) gives the possibility to fixe the constant of oscillator 
$\left(\frac{4}{3}V^3_0 \right)^{1/3}$, if  take the experimental
value  $F_{\pi}= 93 MeV$, then
\ben\left(\frac{4}{3}V^3_0\right)^{1/3}=F_{\pi} L_1(0) = 542 MeV.\een

But such estimate was made in chiral limit. More full picture can be 
obtained by calculation of the leptonic decay constant $F_\pi$ by (11)
with helping of the founding eigenvalue and function of the equation (10). 
From the experimental data it is known
that correlation $M_{\pi}/F_{\pi} \approx 1.5$ that allows to fixe the physical 
parameters of the model (the current quark mass $(m^0)$, the constant of 
interaction potential $\left(\frac{4}{3}V^3_0\right)^{1/3}$ and $\as$).

On fig.2 the dependence of the correlation  $M/F$ on the Coulomb potential
interaction constant $\as$ which is compared with experimental correlation
$M_{\pi}/F_{\pi} \approx 1.5$. 

As it is seen from the figure that there is an agreement between the theory 
and experimental data at the point $\as=0.56$.
 
\bc 5. Conclusion \ec

The quark-antiquark interaction potential chosen in the form
of oscillator and coulomb terms which allows to calculate masses spectra of
the pseudoscalar mesons 
	
In contrast with many different approaches widely developing
in literature, we used the interaction potential must qualitatively
describing the spectrum of mesons without introducing additional physically
meaningless parameters. As repeatedly noted we used in our developed model
only physical parameters namely $q\bar q$- potential parameters ($V_{0}, \as$)
and the current quark masses ($m^0$). The current quark masses are considered 
as $m_{u}=m_{d}$, though our model allow to consider $m_{u}\neq m_{d}$.
 	
The criteria characterizing qualitative description of spectra
 and wave mesons functions is the leptonic decay constant of pion $F_\pi$
 we showed that $F_\pi$ (as opposite value of wave function
$L_{1}(0)$) play role of scale to calculate wave functions and masses
spectra of mesons.

In the paper [3] it was calculated $F_{\pi}$. These values
$F_{\pi}$ were significantly reduced with respect to  experimental
data. In the ref. [11] the given value of constant $F_{\pi}$
approached to the experimental data, the calculation was made on 
changing the oscillator potential parameter
$\left(\frac{4}{3}V^3_0\right)^{1/3} \approx 750 MeV$. Some authors [11]
by means of moving the scale toward higher values attempt to account
for the reason of reduced values of the leptonic decay constant. In our
view, the true reason of difficult calculations $F_{\pi}$ is namely
in the choice of the quark antiquark interaction potential. Our
approach shows the most nicely real picture of pion without adding
some fitting parameters of the model.

The masses spectrum calculated  and the leptonic decay
constant are used for the fixing the coulomb potential parameter
$\as$ by comparing with the experimental ratio  $M_\pi/F_\pi$.
We got $\as=0.56$. Choosing scale or the value of the oscillator potential 
$\left(\frac{4}{3}V^3_0\right)^{1/3} =	515 Β'$, we got  $ F_\pi= 90 MeV$.

In conclusion we may add that the spectra of mesons described by the 
potential model	(3), (4) and (8), and it based on the solution of
the combined equations of Bethe-Salpeter and Schwinger-Dyson,
where the interaction potential chosen in the form of sum of oscillator
and coulomb terms which gives good agreement with the experimental data. 
We wish to note in this approach that the constant of coulomb interaction
potential lies in the field $0.5 \leq \as < 1$.

\end{document}